\documentclass[twocolumn,prl,showpacs,floatfix,altaffilletter]{revtex4}
\usepackage{graphicx}
\pdfoutput=1
\usepackage{bm}
\usepackage{amsmath}
\usepackage{amsfonts}
\usepackage{amssymb}
\begin{document}
\title{Thermopower as sensitive probe of electronic nematicity  in iron pnictides}
\author{Shuai Jiang$^{1,2}$, H.S. Jeevan$^1$, Jinkui Dong$^1$, and P. Gegenwart$^1$}
\affiliation{$^{1}$I. Physikalisches Institut, Georg-August-Universit\"at G\"ottingen, Germany\\$^{2}$1.Physikalisches Institut, Universit\"at Stuttgart, Germany}
\date{\today}
\pacs{74.70.Xa;74.25.F-;74.25.fg;74.40.Kb}

\begin{abstract}
We study the in-plane anisotropy of the thermoelectric power and electrical resistivity on detwinned single crystals of isovalent substituted  EuFe$_{2}$(As$_{1-x}$P$_{x}$)$_2$. Compared to the resistivity anisotropy the thermopower anisotropy is more pronounced and clearly visible already at temperatures much above the structural and magnetic phase transitions. Most remarkably, the thermopower anisotropy changes sign below the structural transition. This is associated with the interplay of two contributions due to anisotropic scattering and orbital polarization, which dominate at high- and low-temperatures, respectively.
\end{abstract}

\maketitle
Electronic states with broken rotational symmetry driven by electronic correlations rather than the anisotropy of the underlying crystal lattice have recently attracted considerable attention~\cite{Kivelson,Borzi,Daou,Okazaki,Stingl}. The iron-pnictide superconductors provide a new way to explore the relation of superconductivity (SC) and electronic nematicity. The AFe$_2$As$_2$ (A=Ba, Sr, Ca or Eu) ("122") materials crystallize in a tetragonal structure at high temperatures. Upon cooling through nearby structural ($T_s$) and magnetic ($T_N$) phase transitions, a low-temperature orthorhombic phase is stabilized where the Fe spins point along the (longer) $a$-axis with antiferromagnetic (AF) alignment ~\cite{Huang}. Along the direction of the (shorter) $b$-axis, neighboring spins are coupled ferromagnetically. The orthorhombic lattice distortion results in the formation of twin domains at $T<T_s$. A small uniaxial pressure along one of the orthorhombic in-plane directions is sufficient for detwinning~\cite{Fisher review}. 

Evidence for a pronounced in-plane electronic anisotropy of 122 systems below $T_s$ has been found in the electrical resistivity~\cite{Chu 2010}, optical response to polarized light~\cite{Dusza,PNAS-Optical}, quantum oscillations~\cite{Terashima} and angular resolved photoemission spectroscopy (ARPES)~\cite{PNAS-ARPES}. The large energy separation of two orthogonal bands with predominant $d_{xz}$ and $d_{yz}$ character found in ARPES, sketched in the lower inset of Figure 1, indicates an orbital polarization at low temperatures~\cite{PNAS-ARPES}. 

Remarkably, even for temperatures well above $T_s$, uniaxial stress induces a pronounced resistivity anisotropy~\cite{Chu 2010}. Using a piezo device, the resistivity anisotropy in the limit of {\it zero} strain has been detected~\cite{Chu 2012}. Indeed, this "nematic susceptibility" diverges in the tetragonal state upon cooling from high $T$ down to $T_s$, even once the latter is suppressed towards $T\rightarrow 0$ by doping. Importantly, electronic nematicity above $T_s$ has also been confirmed on micro crystals with presumed unbalanced twin-domain volumes by magnetic torque measurements~\cite{Matsuda}. 

The origin of the resistivity anisotropy is controversially discussed. In one scenario, it is related to the orbital polarization, even at temperatures above $T_s$~\cite{Chen,Valenzuela}. An alternative scenario has been proposed in~\cite{Schmalian}. The columnar AF ground state of iron-pnictides has a discrete Ising-type symmetry, related to stripes of parallel spins along one of the in-plane axis. Consequently, both the spin rotation and the Ising nematic symmetry are broken below $T_N$ while the state at $T_N<T<T_s$ is characterized by the broken Ising-nematic symmetry only, with nematic spin fluctuations persisting above $T_s$~\cite{Schmalian-Fernandez}. As illustrated in the upper sketch of Fig.~1, the non-zero nematic susceptibility in the electrical resistivity is then caused by the anisotropic scattering of electrons near hot spots of the Fermi surface,  connecting electron and hole pockets (in blue and red respectively, the latter one has elliptical shape due to doping and is shifted by the critical wavevector $(\pi,0)$).

\begin{figure}
\centering
\includegraphics[width=1\columnwidth]{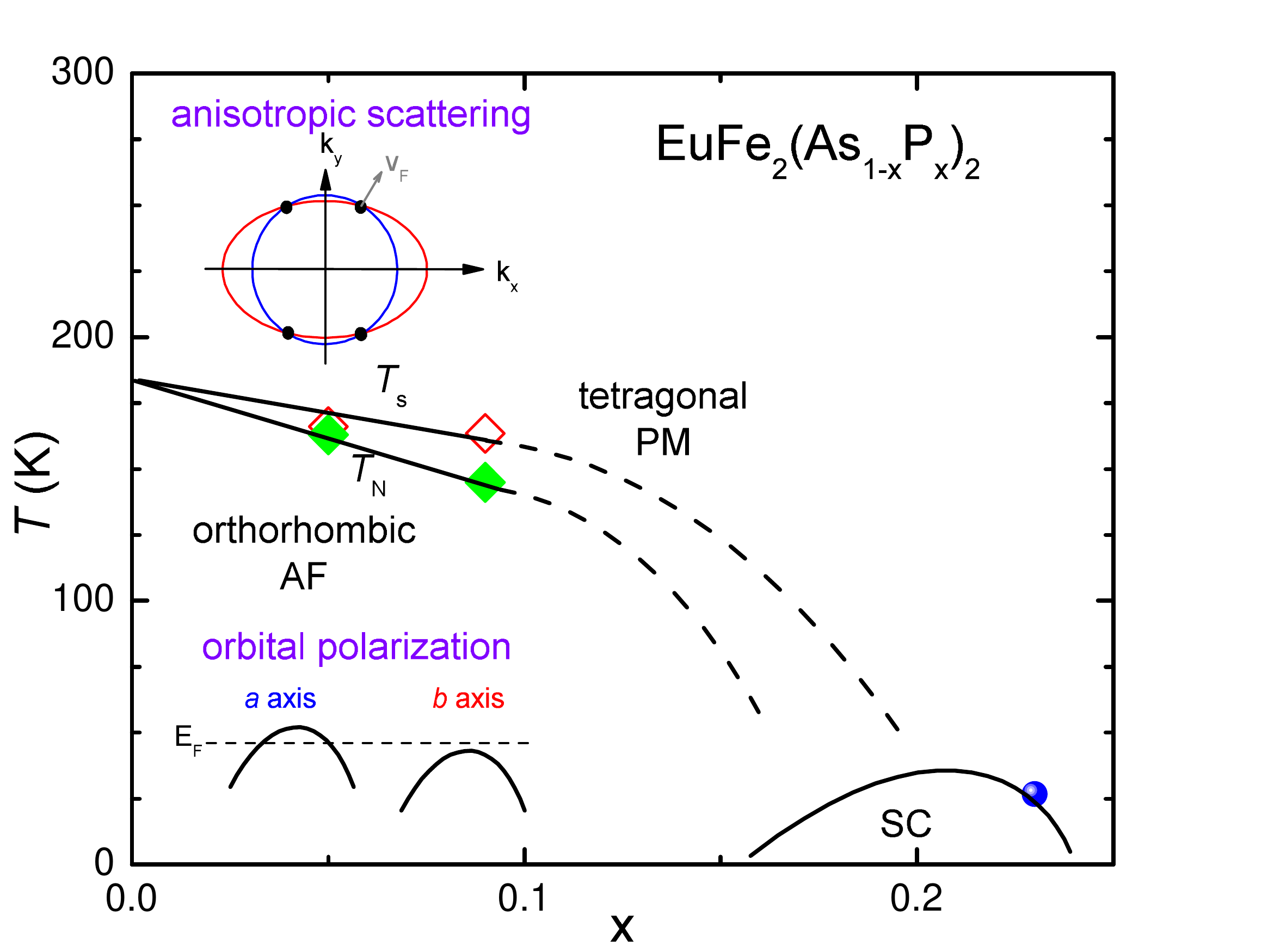}
\caption{\label{fig1}(color online). Schematic phase diagram of EuFe$_{2}$(As$_{1-x}$P$_{x}$)$_2$, based on \cite{Jeevan-prb,Tokiwa}. The ordering temperatures of Eu$^{2+}$ magnetic moments have been omitted for clarity, as they are not important in the context of this study. Red, green and blue symbols denote structural ($T_s$), antiferromagnetic ($T_N$) and superconducting (SC) phase transitions shown in Figs. 2 and 5. The upper and lower cartoons illustrate the anisotropic scattering~\cite{Schmalian-Fernandez} and orbital polarization~\cite{PNAS-ARPES} scenario for the in-plane resistivity behavior, respectively. See text.}
\end{figure}

Below, we establish the thermoelectric power (TEP) as new and particular sensitive probe of electronic nematicity in iron pnictides. Two distinct contributions from anisotropic scattering in the paramagnetic and orbital polarization in the AF state are deduced. We focus on the isovalent substituted system EuFe$_2$(As$_{1-x}$P$_x$)$_2$, for which we have grown large and high-quality single crystals, which previously have been thoroughly characterized by bulk properties, ARPES and optical conductivity~\cite{Jeevan-prb,Tokiwa,Thirupathaiah-prb,Zapf-prb,Wu-prb,Jannis-prb}. The local Eu$^{2+}$ magnetic moments order below about 20~K~\cite{Xiao-EuFe2As2} and have negligible influence on the electronic properties of the system. Similar as for BaFe$_2$(As$_{1-x}$P$_x$)$_2$, the partial substitution of As by isoelectronic smaller P induces a chemical pressure and suppresses the structural and spin-density-wave transitions found for undoped EuFe$_2$As$_2$ near 190~K. ARPES has revealed a non-rigid-band-like change of the electronic structure with P substitution~\cite{Thirupathaiah-prb}. Roughly, the hole-Fermi surfaces become more three-dimensional, thereby weakening the nesting conditions, whereas the size of the electron pocket near the $K$ point slightly increases. The phase diagram is schematically depicted in the main part of Figure 1. As a result of the presence of the Eu$^{2+}$ local magnetic moments whose order below 20~K (not shown here) develops an increasing ferromagnetic component~\cite{Zapf-prb}, bulk SC with $T_{c,max}$ up to 26~K is found only in a narrow composition range and disappears beyond $x=0.23$~\cite{Jeevan-prb,Tokiwa} in contrast to previous reports on polycrystalline samples~\cite{Ren}.

\begin{figure}
{\includegraphics[width=1\columnwidth]{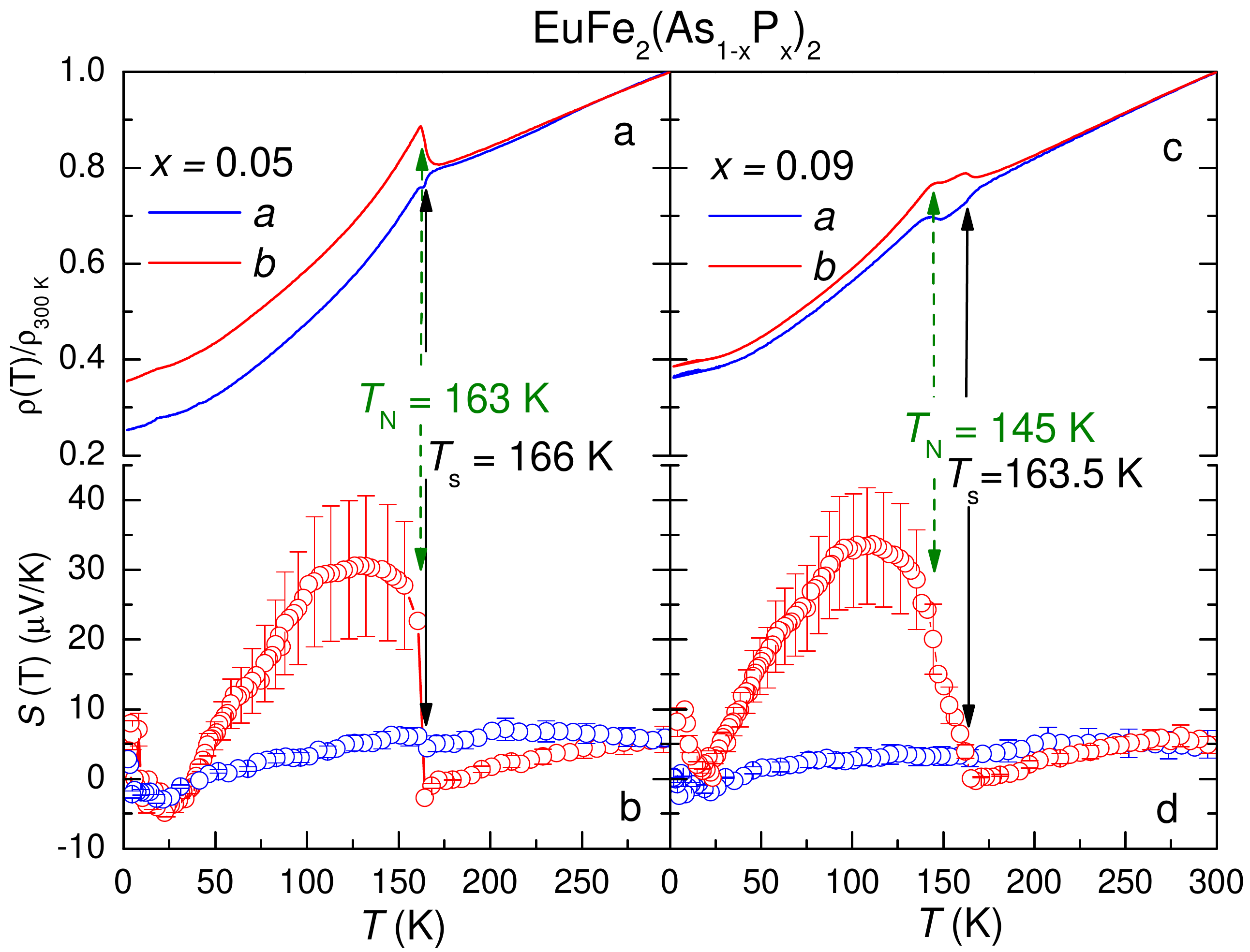}}
{\caption{\label{fig2}(color online). Electrical resistivity (a) and (c) and thermoelectric power (b) and (d) of EuFe$_2$(As$_{1-x}$P$_x$)$_2$ along the orthorhombic a- and b-axis, indicated by blue and red colors, respectively. Solid arrows indicate structural phase transition ($T_s$) as determined by peak position of temperature derivative of resistivity anisotropy (see Fig.~3). The antiferromagnetic transition ($T_N$) is obtained from the (lower) kink of the resistivity data along the b-axis and denoted by the dashed arrows.}}
\end{figure}

We investigate single crystals with compositions $x=0.05$, $0.09$ and $0.23$ whose position in the phase diagram is indicated by the colored symbols in Fig. 1. Note, that $x$ denotes the composition determined by energy dispersive x-ray analysis (with uncertainty $\Delta x\approx 0.01$) rather than the nominal composition. The single crystals used in this study were synthesized and characterized as previously described~\cite{Jeevan-prb}. After orientation, the single crystals were mounted in a uniaxial stress clamp for in-situ detwinning at $T_s$. Details are provided in supplemental material (SM)~\cite{SM}. We used low-temperature polarized light imaging~\cite{Canfield-imaging} in order to prove the single domain state below the structural phase transition. Following previous nomenclature, we use $a$ and $b$ for the longer and shorter in-plane orthorhombic axes, respectively. For TEP measurements, the heat flow between the sample and the clamp has been minimized by using thin plates of Mica for thermal insulation~\cite{SM}. The different samples have also been measured outside the clamp in suspended configuration. For a sample with negligible in-plane anisotropy ($x=0.23$), we could use these measurements to determine a small temperature dependent background correction being proportional to the sample's surface attached to the pressure clamp, which has subsequently been subtracted from the raw data for $x=0.05$ and $x=0.09$~\cite{SM}.

\begin{figure}
\centering
\includegraphics[width=1\columnwidth]{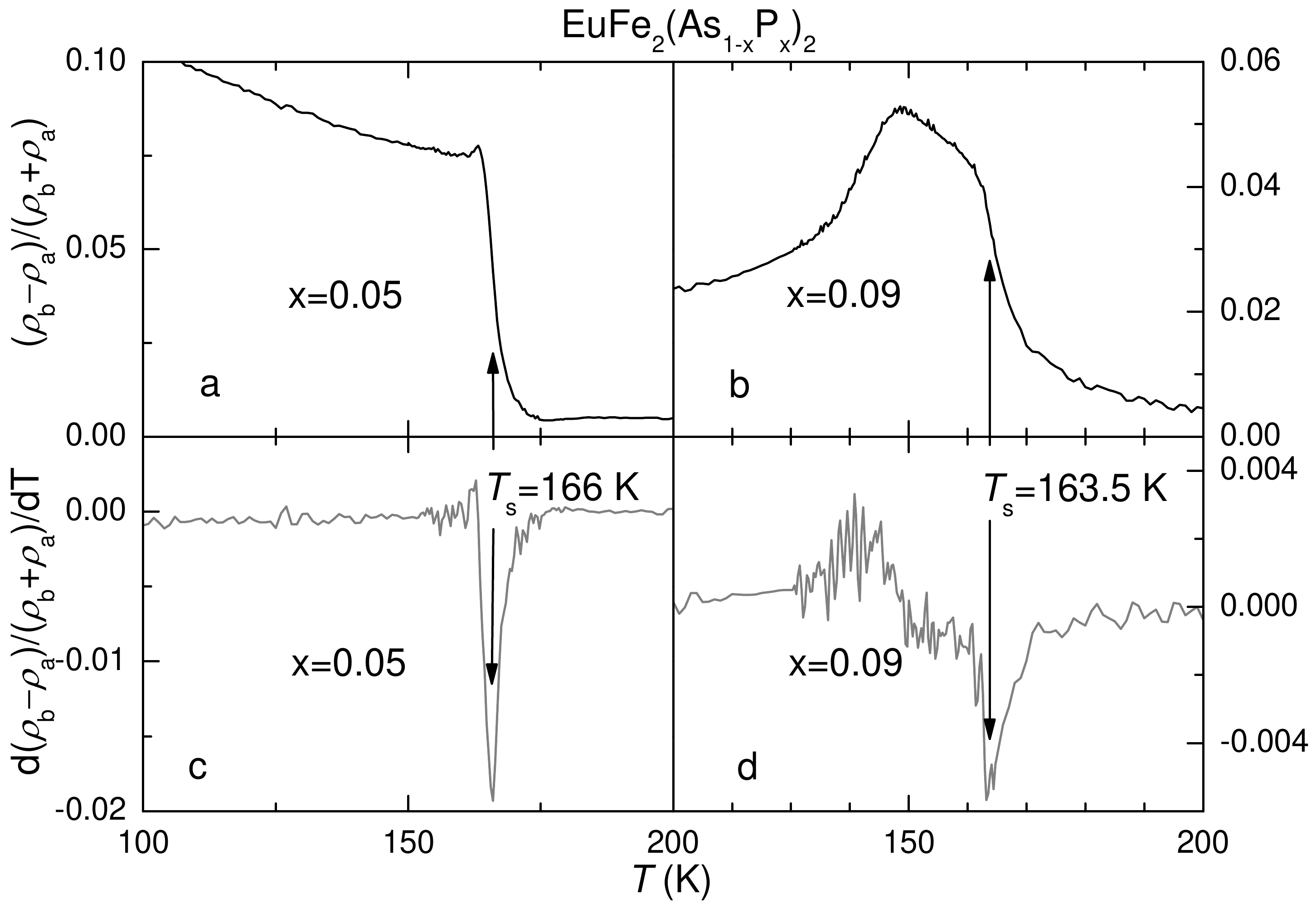}
\caption{\label{fig2}Temperature dependence of the normalized electrical resistivity anisotropy $(\rho_b-\rho_a)/(\rho_b+\rho_a)$ for EuFe$_2$(As$_{0.95}$P$_{0.05}$)$_2$ (a) and EuFe$_2$(As$_{0.91}$P$_{0.09}$)$_2$ (b), as well as respective temperature derivatives (c) and (d). Solid arrows indicate $T_s$ as determined from minima in (c) and (d).}
\end{figure}

We first focus on results on two "underdoped" $x=0.05$ and $x=0.09$ samples, and discuss their resistivity anisotropy above and below the structural and magnetic transitions. As shown in Figure 2, parts a and c, the electrical resistivity along the a- and b-axis ($\rho_a$ and $\rho_b$) displays an almost linear temperature dependence upon cooling from room temperature. However, a careful inspection of the relative difference $\Delta\rho = \frac{{\rho}_b - {\rho}_a}{{\rho}_b + {\rho}_a}$ reveals an anisotropy in the percentage range already in this temperature regime (see Figure 3). We use the sharp peak in the temperature derivative $d\Delta\rho/dT$ to define the position of $T_s$, since at this temperature, the maximal increase of anisotropy is found. As shown in SM (inset of Fig. S3), we also find clear peaks in the temperature derivatives of $\rho_a(T)$ and $\rho_b(T)$ at the same temperature. The AF transition occurs at slightly lower temperatures and gives rise to a further anomaly in the resistivity, which is most prominent along the $b$-axis. Note, that the separation between $T_s$ and $T_N$ increases with increasing $x$~\cite{Tokiwa}. At low temperatures, the resistivity along the $b$-axis is distinctly higher than that along the $a$-axis, similar as found in "electron doped" Ba(Fe$_{1-x}$Co$_x$)$_2$As$_2$~\cite{Chu 2010} and Eu(Fe$_{1-x}$Co$_x$)$_2$As$_2$~\cite{ChenXH-PRL}.

\begin{figure}
\centering
\includegraphics[width=1\columnwidth]{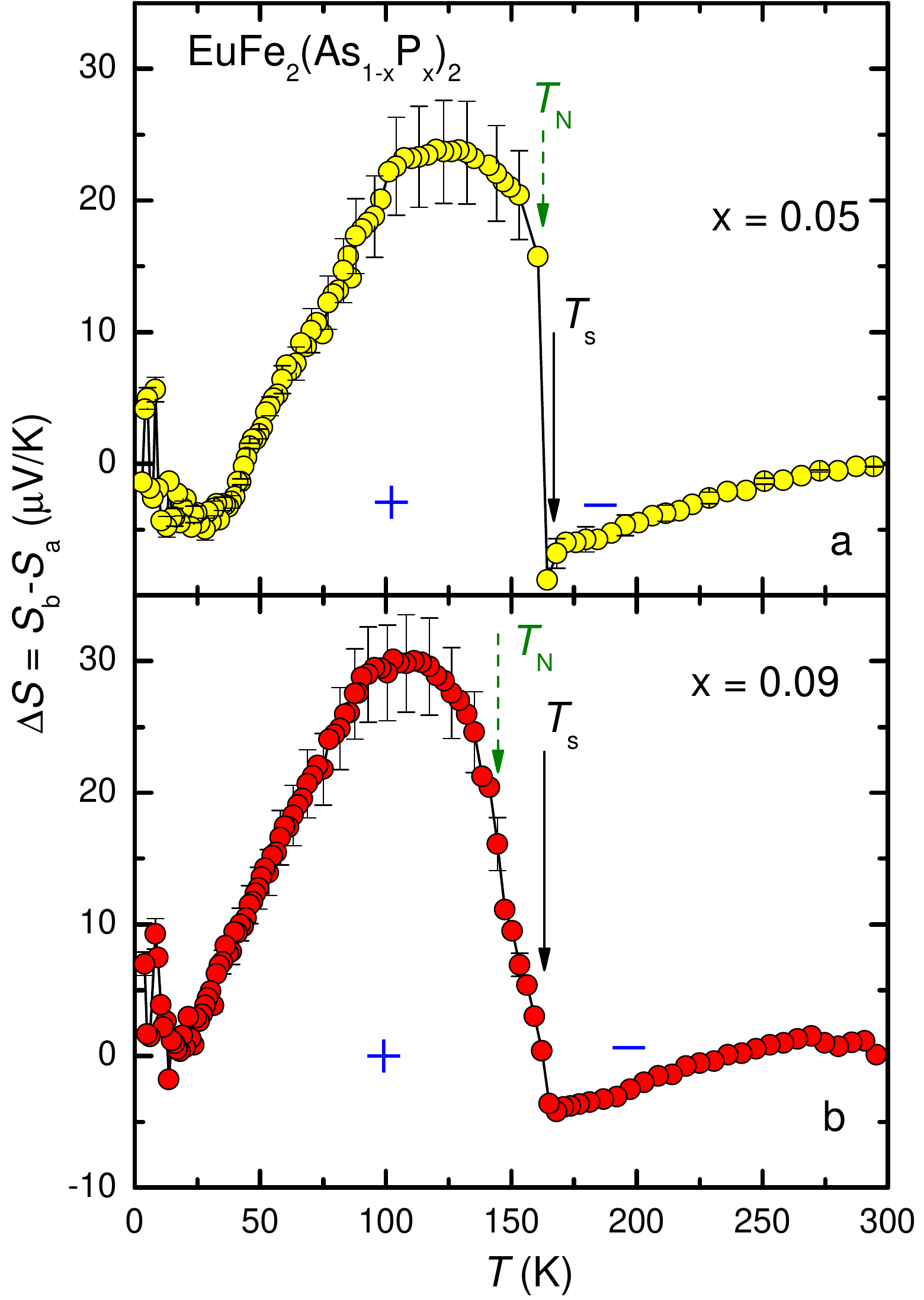}
\caption{\label{fig4}(color online). Temperature dependence of the TEP anisotropy $\Delta S=S_b(T)-S_a(T)$ for EuFe$_2$(As$_{0.95}$P$_{0.05}$)$_2$ (a) and EuFe$_2$(As$_{0.91}$P$_{0.09}$)$_2$ (b). Solid and dashed arrows indicate $T_s$ and $T_N$, respectively, as determined from the electrical resistivity (cf. Fig. 1 and resistivity anisotropy, Fig.~3). Plus (+) and minus (-) symbols indicate sign of thermopower anisotropy.}
\end{figure}

We now turn to the respective anisotropy found in the TEP shown in Fig. 2b and d. As indicated by the solid ($T_s$) and dashed ($T_N$) lines, sharp and well defined signatures, in particular for the data along the $b$-axis, are resolved at the phase transitions. The signature along $a$ is small but clearly visible also in the raw data~\cite{SM}. At high temperature, $S_b(T)$ decreases with decreasing $T$. At $T_s$, a sharp increase of $S_b$ is found and a further change in slope at $T_N$, which is most prominently seen for the $x=0.09$ sample. After passing a maximum around 100~K, $S_b(T)$ then decreases and displays a broadened minimum near 25~K. We have found similar overall behavior in our previous investigation on twinned single crystals of similar compositions~\cite{Jannis-prb}, indicating a dominating contribution from $S_b$ in this latter case. The low-temperature minimum near 25~K has also been found for the hole-doped system Eu$_{1-x}$K$_x$Fe$_2$As$_2$ and been ascribed to a negative phonon drag contribution~\cite{Jannis-prb}.

In order to discuss the anisotropy of the TEP, we calculate the difference ${\Delta S(T)=S_b(T)-S_a(T)}$ between the data along the $b$- and $a$-axis and analyze its temperature dependenc (for a plot of the normalized difference, see SM~\cite{SM}). As shown in Figure 4, $\Delta S(T)$ changes sign at $T_s$: while it is negative at high temperature, a sharp rise sets in at $T_s$, resulting in positive values at least below $T_N$. This is our most important observation and indicates a distinct difference to the electrical resistivity anisotropy, which does not change sign. Furthermore, the normalized anisotropy of the TEP~\cite{SM} is far more pronounced than that of the electrical resistivity, demonstrating the great sensitivity of this property to electronic nematicity.

For discussing the observed anisotropy of the TEP, we start with the Mott expression, calculated from the linearized Boltzmann equation in the degenerate limit $k_BT\ll E_F$

\begin{equation}
\emph{S} = \frac{{{\pi}^2}{k^2_B}\emph{T}}{3e}\frac{{\partial}\ln{\sigma}(E)}{{\partial}E}{\bigg\vert}{_{E_F}},
\end{equation}

which relates the TEP to the logarithmic energy derivative of the electrical (dc) conductivity $\sigma$ at the Fermi energy. Assuming a simple relation $\sigma\propto l  S_F$, where $l$ denotes the mean free path and $S_F$ the Fermi surface area, reveals 

\begin{equation}
\emph{S} = \frac{{{\pi}^2}{k^2_B}\emph{T}}{3e}( \frac{\partial\ln\l}{\partial E}{\bigg\vert}{_{E_F}}+\frac{\partial\ln S_F}{\partial E}{\bigg\vert}{_{E_F}}).
\end{equation}

The two contributions to the TEP are arising from the scattering, determining the mean-free path, and from the band structure. The in-plane anisotropy of the TEP under a small uniaxial pressure is therefore determined by the sum of the terms reflecting the anisotropies of the mean free path and Fermi surface, respectively. The former will be affected by anisotropic scattering due to magnetic fluctuations while the latter is induced by orbital polarization. The Fermi surface reconstruction below $T_N$ may also add to the latter contribution. However, mean-field calculations have revealed that the observed band anisotropy could not be explained by magnetic order alone and requires orbital polarization~\cite{Lv}. Given that the TEP anisotropy depends on the anisotropy of $l$ and $S_F$, the question arises which term dominates in which temperature regime.

\begin{figure}
\centering
\includegraphics[width=1\columnwidth]{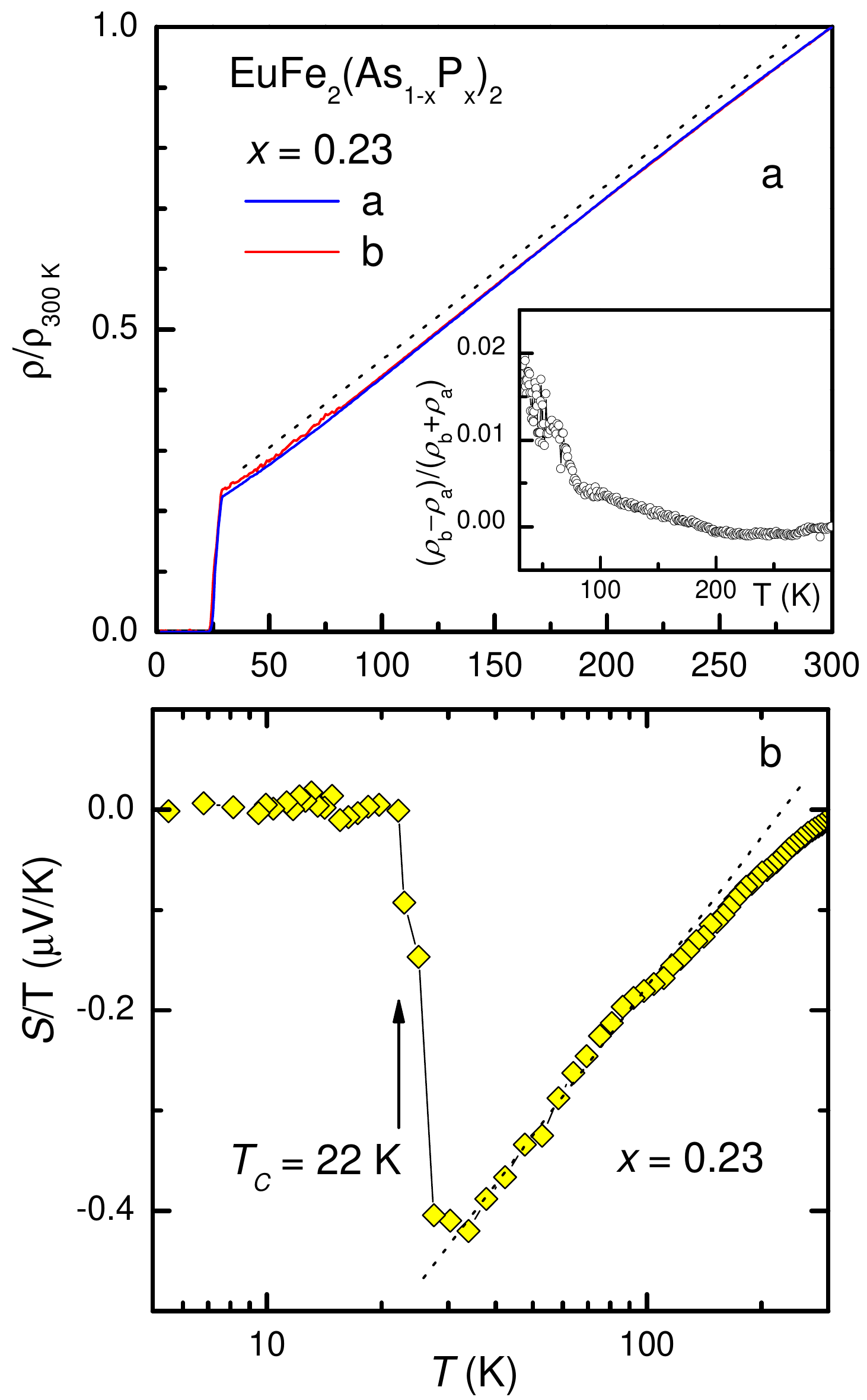}
\caption{\label{fig5}(color online). Temperature dependence of the electrical resistivity (a) and TEP (b) for EuFe$_{2}$(As$_{0.77}$P$_{0.23}$)$_2$ (a and b denote in-plane  tetragonal (110) directions perpendicular and parallel to the applied uniaxial pressure, which in the orthorhombic notation would correspond to the respective main axes, however, there is no orthorhombic phase in this sample). For (b) data along the tetragonal (110) axis without uniaxial pressure are shown~\cite{SM}. The inset in (a) displays the resistivity anisotropy in the normal state. Dashed lines are intended as guide to the eye and arrow indicates transition into the SC state.}
\end{figure}

At temperatures above $T_s$, the orbital polarization is weak and respectively the strain dependence of the Fermi surface is expected to be small. On the other hand, nematic spin fluctuations will result in a pronounced anisotropy of the mean-free path. Since the resistivity along the (shorter) $b$-direction is higher than along $a$, the uniaxial pressure dependence of the mean free path is negative along $b$ and positive along $a$ which naturally explains $\Delta S=S_b-S_a<0$. This contribution is expected to decrease below $T_N$ once the spin fluctuations are suppressed.

The dramatic increase of $\Delta S(T)$ at $T\leq T_s$ likely reflects the dominance of the orbital polarization, leading to a significant shift of the density of states below the Fermi energy along the $b$-axis (cf. lower sketch in Fig.~1). Respectively, a large increase of the TEP is found along this direction, resulting in $\Delta S>0$. 

Within this picture the sign change of TEP points to the competition of two contributions due to anisotropic fluctuations, dominating at high temperatures and orbital polarization, dominating at low temperatures. If, as proposed in \cite{Chen}, the orbital polarization would be dominating already at high temperatures and be the driver of the anisotropy in the initial strain dependence of the resistivity, a sign change of TEP would not arise. Therefore, our data support the view \cite{Schmalian-Fernandez}, that nematicity at high temperatures results from anisotropic magnetic scattering.

At last, we focus on a composition $x=0.23$ in the slightly "overdoped" regime of the phase diagram of EuFe$_2$(As$_{1-x}$P$_x$)$_2$  (cf. Fig.~1). A previous ARPES study on the system has revealed a Lifshitz transition near $x=0.21$ at which the inner hole pocket along $\Gamma-\emph{Z}$ disappears~\cite{Thirupathaiah-prb}. The thermopower $S(x)$, being very sensitive to the Fermi surface, has detected a non-monotonic evolution at this concentration at all investigated temperatures~\cite{Jannis-prb}, similar as found near Lifshitz transitions in electron doped Ba(Fe$_{1-x}$Co$_x$)$_2$As$_2$~\cite{Lifshitz}. The structural transition is completely suppressed for $x=0.23$ in accordance with the phase diagram. Indeed, as shown in Figure 5a, almost no in-plane anisotropy of the electrical resistivity could be detected. No reliable anisotropy of the TEP could be detected either~\cite{SM} and the data agree within the experimental error with those measured previously without uniaxial pressure clamp~\cite{Jannis-prb}. Therefore, Fig. 5b only includes data measured along one direction.

Similar as found for the related system BaFe$_2$(As$_{1-x}$P$_x$)$_2$~\cite{Kasahara-prb}, an almost linear temperature dependence of the normal-state electrical resistivity is found from $T_c$ up to room temperature (cf. dotted line in Fig.~5a). This non-Fermi liquid behavior is further corroborated by the logarithmically divergent coefficient of the TEP, $S/T\propto \log T$ shown in Fig.~5b. These temperature dependences would be compatible with two-dimensional AF quantum critical fluctuations~\cite{Paul}. The same conclusion has also been drawn from NMR measurements on BaFe$_2$(As$_{1-x}$P$_x$)$_2$~\cite{Nakai} and a sharp peak of the zero-temperature SC penetration depth for this latter system suggests a quantum critical point located in the SC state at optimum substitution~\cite{Hashimoto}.

We have used the TEP to investigate electronic nematicity in the isovalent substituted iron-pnictide EuFe$_2$(As$_{1-x}$P$_x$)$_2$ on detwinned single crystals using a uniaxial-pressure technique. It turns out, that TEP is a very sensitive probe and displays a pronounced anisotropy for low substitution $x$. Remarkably, this anisotropy changes sign upon cooling from above to below the structural phase transition. We propose two contributions of opposite sign arising from anisotropic scattering dominating at $T>T_s$ and orbital polarization for $T<T_s$. For the future, it will be interesting to perform similar experiments on hole doped 122 pnictides, for which previous electrical resistivity measurements have found an opposite sign of the anisotropy~\cite{Blomberg-sign}.

\begin{acknowledgments}

We thank R.M. Fernandes, W. Ku, J. Schmalian, and Z.W. Zhu for informative discussions, J. Maiwald for help with the TTO option of the PPMS and  J. Norpoth and C. Jooss for providing access to a low-temperature polarized light microscope. S.J. and J.D. acknowledge support from the Alexander von Humboldt Foundation. This work is supported by the DFG through SPP 1458.
\end{acknowledgments}


\vspace{1cm}

\textbf{\em{Supplemental Material}}


%
%
%

\section{\bf S1. Sample preparation and experimental setup}

EuFe$_{2}$(As$_{1-x}$P$_{x}$)$_2$ single crystals were grown by the Bridgman method, whose detailed procedure is described in~\cite{Jeevan-prb_s}. Plate-like single crystals were oriented by Laue diffraction at room temperature, as shown in Fig. S1(a). Subsequently, they were cut to rectangular (almost square) shape of typically $\sim 2$~mm in-plane dimensions with the two long axis aligned along the two equivalent tetragonal (110) directions. Afterwards, the single crystals were quartz-glass sealed under Ar gas and gently annealed at $500^\circ$C for 12~h, in order to reduce possible imperfections and to improve the resistivity ratio~\cite{PNAS-Uchida_s}.

\begin{figure}[t!]
\includegraphics[width=1\columnwidth]{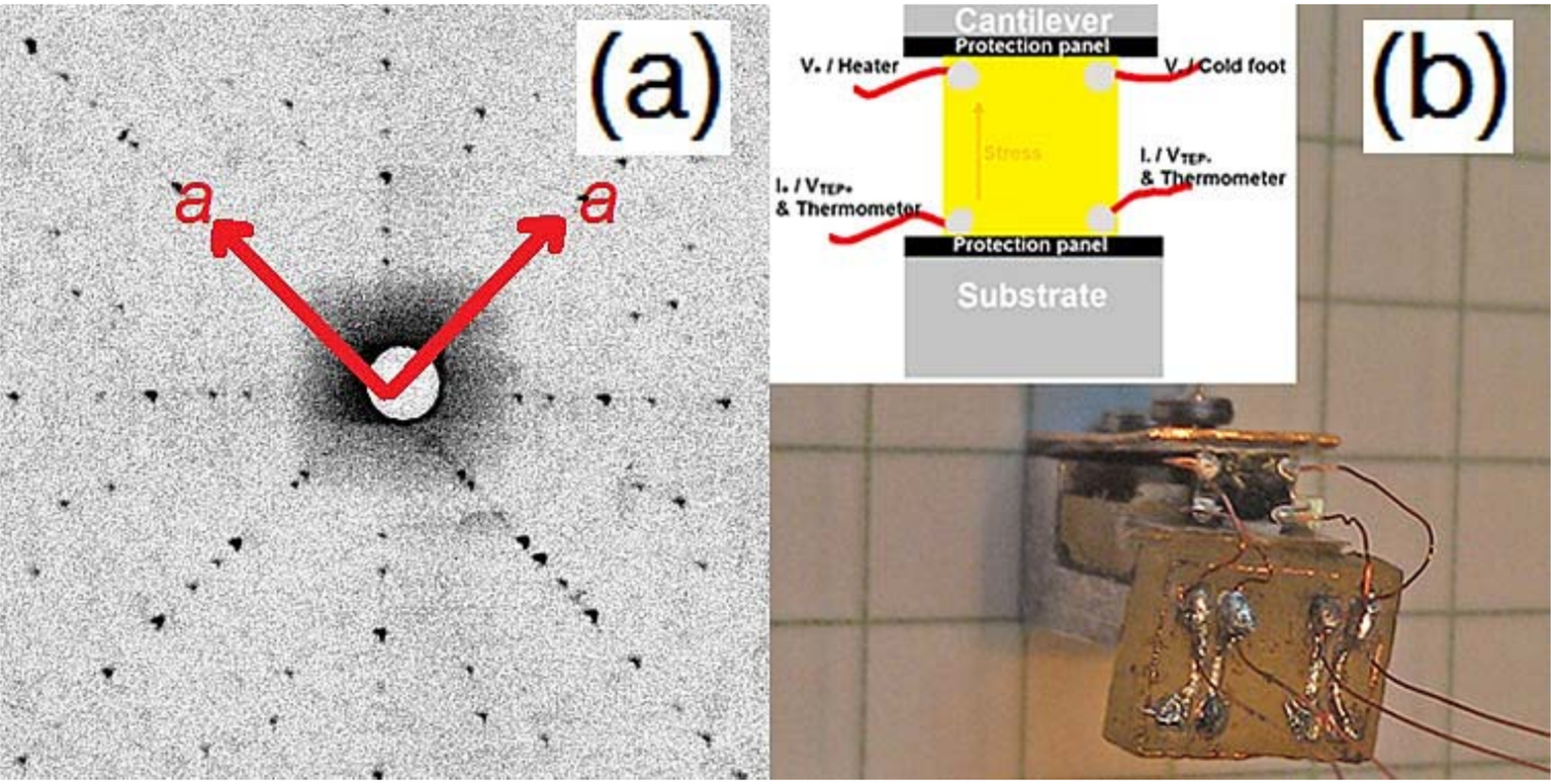}
\renewcommand{\thefigure}{S1}
\caption{(a) Laue pattern of EuFe$_2$(As$_{0.95}$P$_{0.05}$)$_2$ at room temperature, i.e. within the tetragonal state. The crystals have been cut along the tetragonal (110) directions labeled $a$. (b) Photograph of the pressure clamp before mounting on the PPMS puck. The wires soldered to the plastic holder were suspended and shortened before the measurements. The inset of (b) sketches the configuration for electrical resistivity and thermoelectric power (TEP) measurements along one direction (for the perpendicular direction, connections are rotated by $90^\circ$). The protection panel is made from Mica, while the substrate material is Torlon.}
\end{figure}

\begin{figure}
\centering
\includegraphics[width=1\columnwidth]{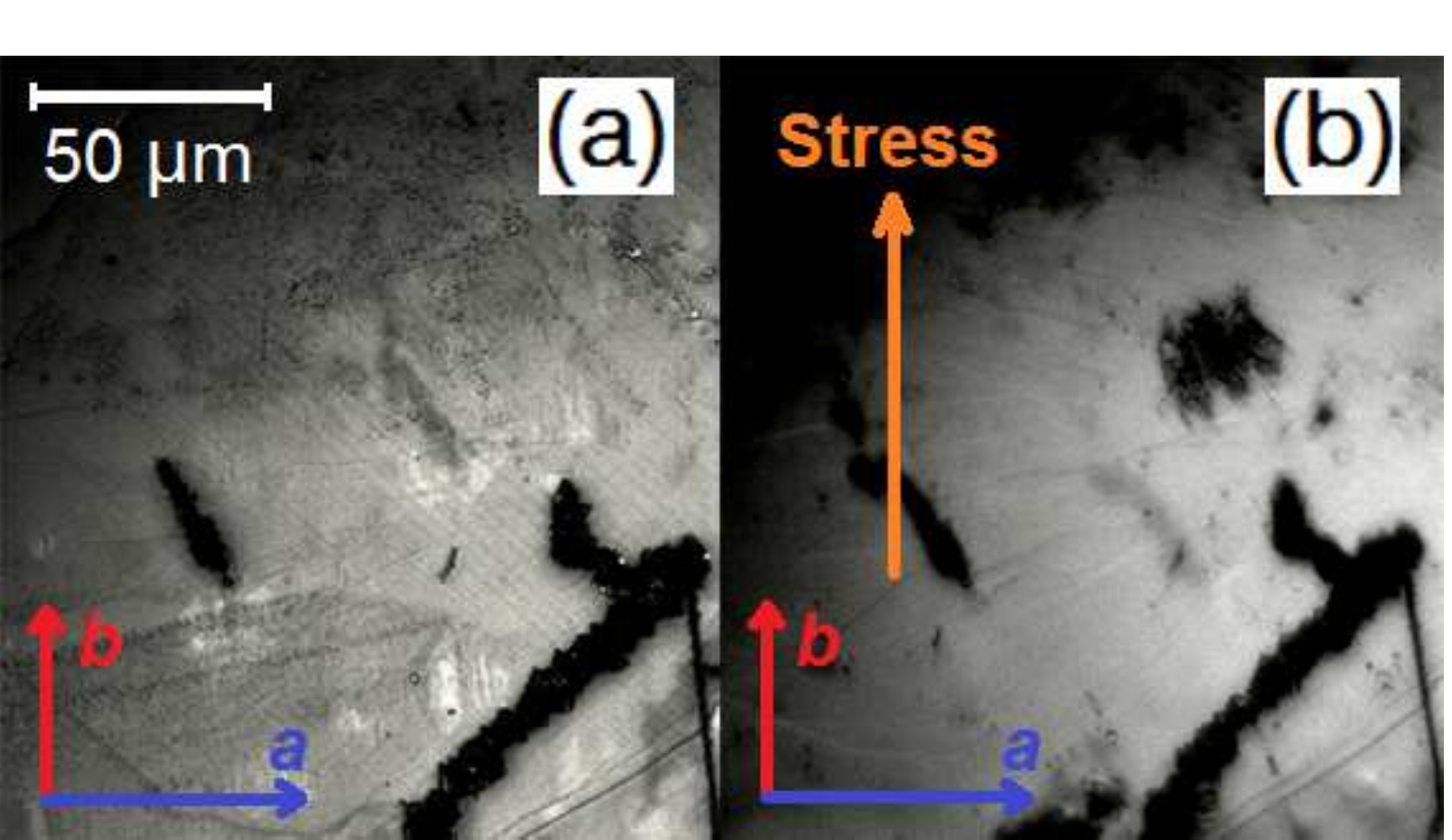}
\renewcommand{\thefigure}{S2}
\caption{Optical imaging of EuFe$_2$(As$_{0.95}$P$_{0.05}$)$_2$ at 30 K without uniaxial stress (a) and after application of uniaxial stress of about 6~MPa along the direction indicated by the orrange arrow (b). In the orthorhombic state, the shorter \emph{b} axis is aligned parallel to the uniaxial pressure~\cite{ChenXH-PRL_s, Chu-PRB_s, review_s}. Note that the two images cover the same sample area (the horizontal label "7" has been scratched before, to find back the same position on the sample surface).}
\end{figure}

For detwinning, we constructed a small uniaxial pressure clamp shown in Fig. S1b, following the design of~\cite{Science-Chu_s}. Uniaxial stress is applied to the sample by the distortion of a CuBe cantilever. Using the Young's modulus of CuBe, the geometrical parameters and the deflection of the cantilever, we can estimate the force/stress which is applied to the mounted single crystal. A bilayer panel of Torlon and Mica for electrical and thermal insulation was pasted on the cantilever and the substrate. This method can protect the single crystal from the electrical short-circuit and the large loss of thermal current in our transport measurements.

Polarized light imaging is an effective way to monitor the difference between twinned and detwinned phases~\cite{Canfield-imaging_s, ChenXH-PRL_s, review_s}. First, the $x=0.05$ and $x=0.09$ single crystals were mounted freely on the stage of a polarized white light microscope. Upon cooling to below the respective structural phase transition temperatures $T_s$, we observed stripes, as shown in Fig. S2a. They are caused by the birefringence between two twinned domains. After mounting the same samples into the clamp for detwinning under a stress of $\sim$ 6 MPa, no more stripes were observed even well below $T_s$ (Fig. S2b). This proves that the single crystal has been detwinned allowing to uncover the in-plane anisotropy of physical properties.

\section{S2. Electrical resistivity measurements}

\begin{figure}
\includegraphics[width=1\columnwidth]{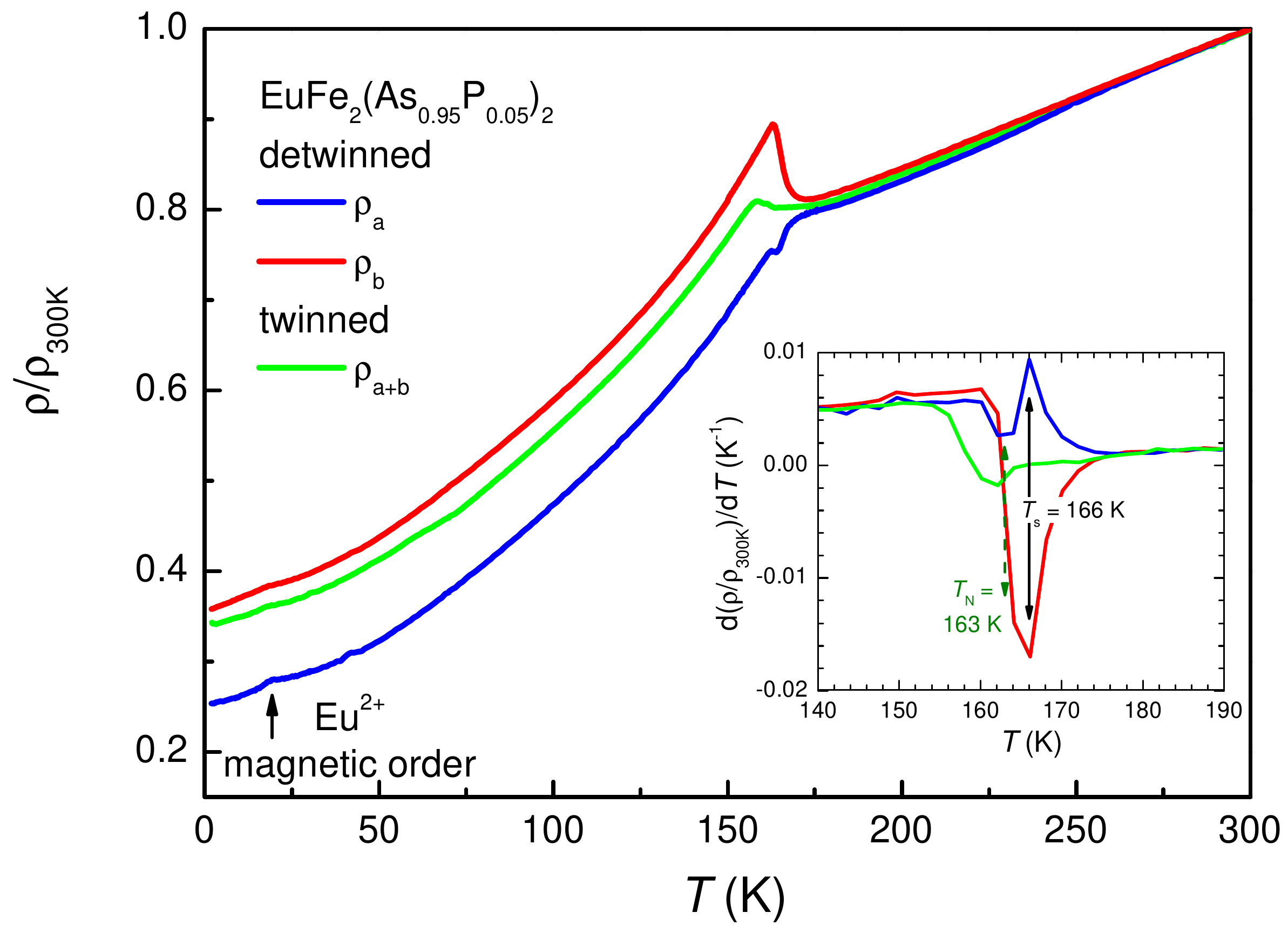}
\renewcommand{\thefigure}{S3}
\caption{Temperature dependence of the in-plane electrical resistivity of EuFe$_2$(As$_{0.95}$P$_{0.05}$)$_2$ with the electrical current along the longer $a$ (blue) and shorter $b$ (red) directions. The green curve shows the in-plane resistivity $\rho_{a+b}$ measured for the twinned sample for comparison. The kink of resistivity data at 19 K is caused by the magnetic ordering of Eu$^{2+}$ moments~\cite{Jeevan-prb_s}. The inset displays the respective temperature derivative of the resistivity data from the main part. The arrows indicate the transition temperatures, determined as discussed in the main part of the manuscript.}
\end{figure}

Subsequent to the demonstration of the detwinning, we made electrical contracts in situ on the crystals mounted in the pressure clamp. Dupont 4929N silver epoxy and 80 $\mu$m diameter copper wires were used and the contacts were arranged in the Montgomery configuration (cf. Fig. S1b). The temperature dependent resistivity measurements were conducted using the standard ac technique in the Quantum Design PPMS. The size of contacts were much smaller than the length of the single crystal. The contact resistances were less than 1 $\Omega$. With the current flowing parallel to the orthorhombic \emph{a} axis and b axis on the detwinned single crystal, $\rho$$_a$ and $\rho$$_b$ were measured at the same time by the Montgomery method. The resistivity of the twinned single crystal, labeled $\rho_{a+b}$, was determined after releasing the uniaxial stress. Fig. S3 displays the resistivity measurements on EuFe$_2$(As$_{0.95}$P$_{0.05}$)$_2$ along the $a$- and $b$-direction (see also Fig. 1a of main text) together with the data from the twinned crystal in the main part and respective temperature derivatives in the inset.The arrows indicate the transition temperatures $T_s$ and $T_N$, determined as described in the main text. It appears that the structural and magnetic transitions of the twinned crystal are slightly lower than those of the detwinned one. We ascribe this to the effect of in-plane uniaxial stress, which is known to increase both transitions~\cite{Wilson_s}.

\section{S3. Thermoelectric power measurements}

\begin{figure}
\centering
\includegraphics[width=1\columnwidth]{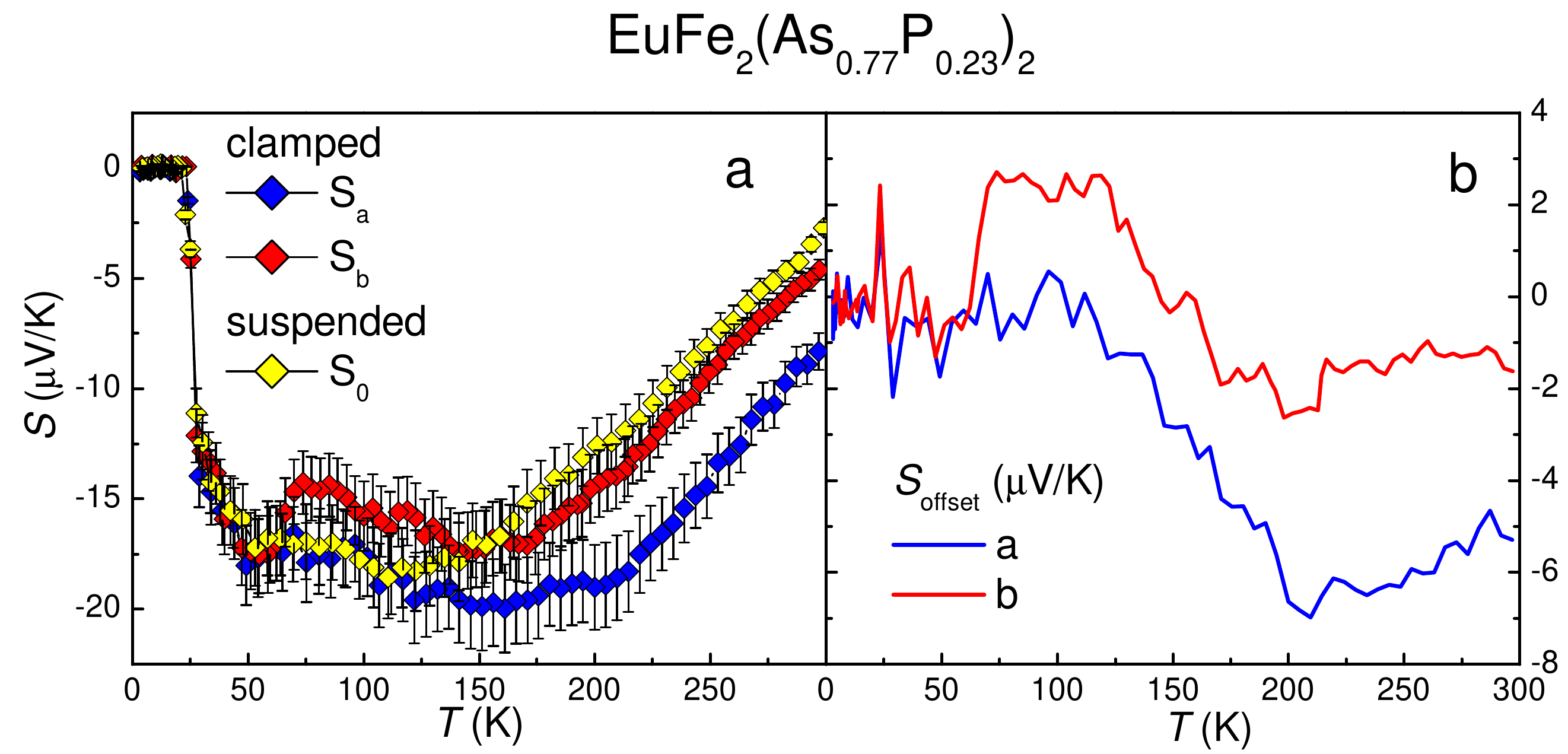}
\renewcommand{\thefigure}{S4}
\caption{(a) Raw data of the TEP measurements on EuFe$_2$(As$_{0.77}$P$_{0.23}$)$_2$ within the uniaxial pressure clamp ($S_a$ and $S_b$) and, using same contacts, for the freely suspended configuration ($S_0$). Error bars are calculated from the ratio of the individual contact sizes  to their separation. (b) Offset TEP due to non negligible heat flow from the sample through the thermal insulation towards the pressure clamp, as determined by $S_{\rm offset,a}=S_a-S_0$ and $S_{\rm offset,b}=S_b-S_0$.}
\end{figure}

Subsequent to the electrical resistivity measurements, we measured the TEP in situ by using the same contacts. The measurement was carried out in the PPMS system equipped with the high-vacuum option. The temperature gradient was obtained using one heater and one cold-foot connection. By switching the shoes connecting to electrodes on the single crystal, the heat flow could be applied along the \emph{a} or \emph{b} directions, as for the electrical resistivity measurements. Note, that the same contacts are used for the detection of the voltage drop $\Delta V$ and the temperature gradient $\Delta T$. Thus, absolute values of the TEP $S=\Delta V/\Delta T$ without correction related to the geometry could be obtained within the Montgomery configuration, in contrast to the case of electrical resistivity. The raw data of the TEP for $x=0.23$ and $x=0.05$ are shown in Figs. S4a and S5a, respectively.

\begin{figure}
\centering
\includegraphics[width=1\columnwidth]{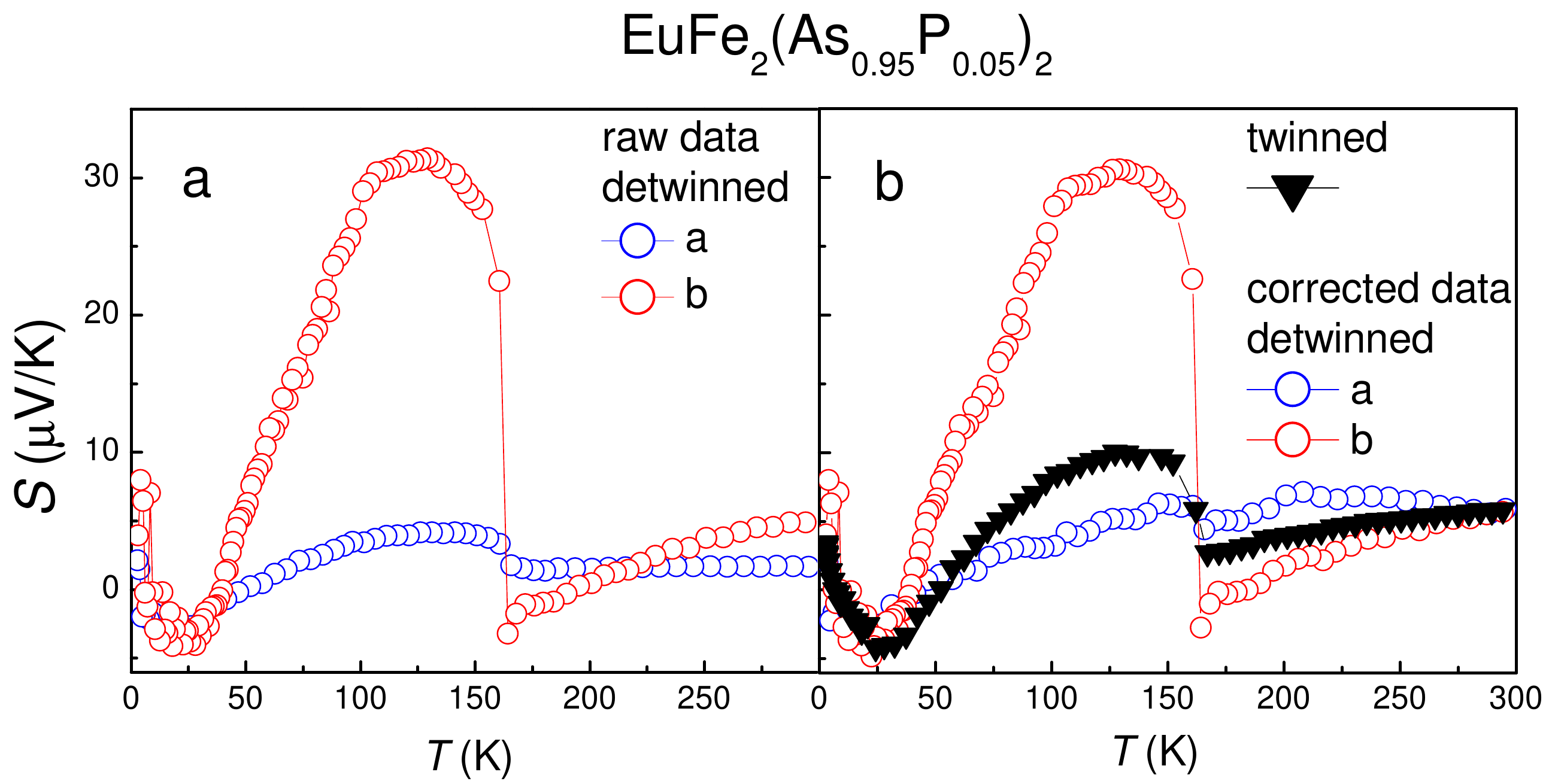}
\renewcommand{\thefigure}{S5}
\caption{(a) Raw data of the TEP  measured within the pressure clamp for EuFe$_2$(As$_{0.95}$P$_{0.05}$)$_2$. The blue and red curves show the Seebeck coefficient along the $a$ and $b$-axis, respectively. (b) Corrected data for the TEP along $a$ and $b$ together with data taken for the freely suspended sample (twinned). For the correction, the offset TEP data of Fig. S4b have been scaled by the respective cross sectional areas between sample and pressure clamp and subsequently subtracted from the raw data.}
\end{figure}

Since for the overdoped $x=0.23$ crystal no anisotropy in the electrical resisitvity is found at room temperature (cf. Fig. 5a of main text), it is highly unlikely, that the observed thermopower anisotropy at room temperature is intrinsic to the sample. We therefore attribute the latter to the (at high temperatures unavoidable) thermal coupling of the sample through the insulator material to the pressure clamp. Although we have used Mica as excellent thermal insulator, its phonon thermal conductivity at high temperatures may result in some heat flow from the heater to the pressure cell. This will modify the drift of phonons in the sample from the warm to the cold end. A slight modification of this drift, in particular at high temperatures, may then influence the TEP as well. In the following, we describe in detail how we compensate for this effect.

\begin{figure}
\centering
\includegraphics[width=1\columnwidth]{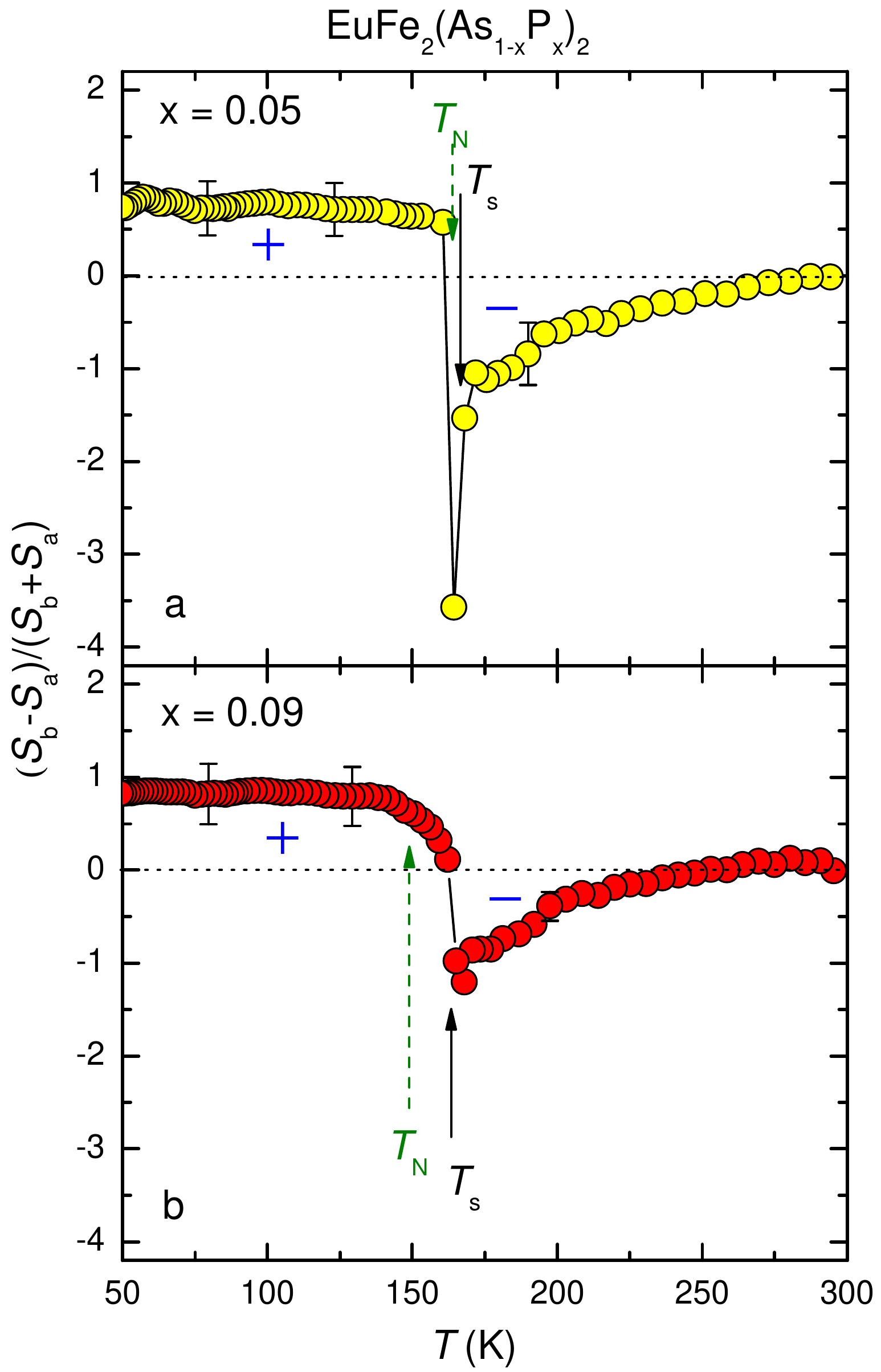}
\renewcommand{\thefigure}{S6}
\caption{(a) Temperature dependence of the normalized TEP anisotropy $(S_b-S_a)/(S_b+S_a)$ for EuFe$_2$(As$_{1-x}$P$_{x}$)$_2$ with $x=0.05$ (a) and $x=0.09$ (b). The blue and red curves show the Seebeck coefficient along the $a$ and $b$-axis, respectively. (b) Solid and dashed arrows indicate $T_s$ and $T_N$, respectively, as determined from the electrical resistivity (main text) while plus and minus symbols indicate the sign of the anisotropy.}
\end{figure}

The comparison with the data measured for the freely suspended sample, denoted $S_0$, using the same contacts as for $S_b$, allows to define a background TEP for the two configurations as
$S_{\rm offset,a}=S_a-S_0$ and $S_{\rm offset,b}=S_b-S_0$ (cf. Fig. S4b). Since this background results from the thermal coupling of the sample to the pressure clamp, it is proportional to the contact area between sample and clamp. We can thus obtain normalized offset contributions by dividing by this contact area (not shown).

For the analysis of the measurements on the other crystals $x=0.05$ and $x=0.09$, we have multiplied the two normalized offset contributions along $a$ and $b$ (from the measurement on $x=0.23$) by the respective cross sectional areas between these latter crystals and the pressure clamp in the respective configurations and subtracted such offset contributions from the raw TEP data.

Fig. S5a displays the raw data for the $x=0.05$ sample obtained within the pressure clamp. Importantly, the main signature at $T_s$, which is the sign change of the TEP anisotropy, is clearly visible already in the raw data. However, these data also display some unexpected difference at room temperature, which is most likely related to the heat flow through the pressure clamp. Subtracting the offset contribution from the $x=0.23$ measurement, scaled with the respective cross sections (see above) indeed yields corrected data which merge at room temperature. In addition, the data measured on a freely suspended sample nicely lie in between the corrected $S_a$ and $S_b$ which convinces us on the reliability of the analysis. Using a background correction determined from data for $x=0.23$, which displays {\it isotropic} TEP behavior, thus allows to correct the data for $x=0.05$ and $x=0.09$ and analyze their {\it anisotropic} behavior. 

Finally, we show in Fig. S6 the normalized anisotropy of the TEP. While the respective normalized anisotropy of the electrical resistivity (Fig. 3a and b of main text) maximally amounts to 0.1, much larger anisotropy is found in the TEP. Within the magnetically ordered state at $T<T_N$ the normalized anisotropy $(S_b-S_a)/(S_b+S_a)$ reaches approximately 1, i.e. the maximal possible value for $S_b>0$, $S_a>0$ and $S_b \gg S_a$. This indicates the enormous sensitivity of the TEP to the anisotropic electronic properties in iron-pnictides.


\end{document}